\documentstyle[12pt]{article}
\begin{document}
\thispagestyle{empty}

\begin{center}
\LARGE \tt \bf{Energy spectra of spinning particles around spinning strings in Einstein-Cartan gravity}
\end{center}

\vspace{2.5cm}

\begin{center} {\large By  L.C. Garcia de Andrade\footnote{Departamento de
F\'{\i}sica Te\'{o}rica - Instituto de F\'{\i}sica - UERJ

Rua S\~{a}o Fco. Xavier 524, Rio de Janeiro, RJ

Maracan\~{a}, CEP:20550-003 , Brasil.}}
\end{center}

\vspace{2.0cm}

\begin{abstract}
Riemannian Killing conserved currents are used to find the energy spectra of classical spinning particles moving around a spinning string in Einstein-Cartan (EC) theory of gravity. It is shown that a continuos spectrum is obtained for planar open orbits and the energy shift of spectrum is obtained for circular planar orbits. A spin-spin effect between the spin of the test particle and the spin of the spinning string contributes to the energy spectrum splitting.
\end{abstract}

\newpage
\pagestyle{myheadings}
\markright{\underline{spinning particles in EC}}
\section{Introduction}

Recently the interest in placing limits to spacetime torsion has led several authors \cite{1,2,3,4,5} to investigate the energy spectra of several metrics in Einstein-Cartan gravity. A commom feature among these authors was to make use of Klein-Gordon scalar particles or Dirac spinning particles \cite{2}. Figueiredo et al \cite{2,3} have used cosmological metrics and found continuous and discrete spectra for the test particles in these spacetimes. On the other hand Lammerzahl  \cite{1} has used the Hughes-Drever experiment to place upper limits on Cartan torsion with hydrogen-like atoms with terretrial laboratory experiments. In this paper we decide to addopt a simpler approach to obtain the spectra of the spinning particles moving around spinning strings and use classical spinning particles instead of quantum particles like electrons and neutrinos. The advantage of using spinning classical particles instead of quantum particles is that we may extend the approach here to the investigation of chaos of spinning particles around cosmic spinning strings in the same way Susuki and Maeda \cite{6} have done  previously in the context of spinning particles around black holes. The paper is organized as follows. In the section 2 we review the mathematical framework of the motion of spinning particles in spacetime backgrounds due to torsion as proposed by Hojman \cite{7} and apply his Killing currents and conserved quantities associated with a spinning particle moving on a spacetime background endowed with torsion to the spinning particle system around a cosmic string and obtain algebraic relations between the spin tensor components and the linear momenta of the spinning particles and the metric of spinning cosmic string in EC gravity as given previously by Soleng \cite{8} and Letelier \cite{9}. In this section three we compute the energy spectrum of spinning particles for open planar orbits around the spinning strings and show that the spectrum obtained is a continuous spectrum. In this same section the analogous computation for a closed circular orbit is undertaken and the energy shift for the spectrum is computed from the higher and lower bound for the energy obtained. We also show that the spin of the particle and the spin polarization density of the particles inside the spinning string yield a spin-spin effect. As a particular case the enrgy shift for the spinless test particles are computed. Spin-orbit effects are also obtained. It is very important to realize that we have used Riemannian conserved currents instead of non-Riemannian Hojmann ones since in this case though the spin density of the spinning string has some effect on the exterior side of the spacetime defect through the matching conditions Cartan torsion vanishes off the string. Spinning cosmic strings and pseudoclassical particles with Klein-Gordon spectra have been also considered by Lorenci et al \cite{10}.
\section{Killing symmetries in non-Riemannian spacetime}
The use of Killing symmetries in General Relativity (GR) have been proved very useful in building the equations of motion of spinning particles in Riemannian spacetimes mainly by Papapetrou \cite{11} and Dixon \cite{12}. More recently Hehl \cite{13} and Hojman \cite{7}  to the case of non-Riemannian spacetime with Cartan geometrical torsion. They are given by the set of equations 
\begin{equation}
\frac{d{x}^{i}}{d{\tau}}=v^{i}
\label{1}
\end{equation}
where $(i,j=0,1,2,3)$ and 
\begin{equation}
\frac{Dp^{i}}{D{\tau}}= {R^{i}}_{jkl}({\Gamma}) v^{j} S^{kl}-{K^{i}}_{kl}v^{k} p^{l}
\label{2}
\end{equation}
\begin{equation}
\frac{DS^{kl}}{D{\tau}}=p^{k}v^{l}-p^{l}v^{k}
\label{3}
\end{equation}
Here ${\tau}$ is the proper time of the orbit , while $S^{kl}$ ,$v^{k}$ and $p^{k}$ are respectively the spin tensor, four-velocity and four-momentum of the spinning particle. The proper time is chosen in such a way that $v^{k}v_{k}=-1$. As in GR chaos the multipoles higher than the monopole and spin dipole are ignored. This approximation is called pole-dipole.
The absolute derivative $\frac{D}{D{\tau}}$ on the RHS of equations (\ref{2}) and (\ref{3}) is the Riemannian derivative with respect to the Christoffel connection. The connection ${\Gamma}$ in the Riemann-Cartan connection. Since $p^{k}$ is no longer parallel to the velocity $v^{l}$ we need a supplementary condition called Dixon condition on the center of mass and given by 
\begin{equation}
p_{j}S^{ij}=0
\label{4}
\end{equation}
Here the mass of the spinning particle ${\mu}$ is given by
\begin{equation}
{m}^{2}= -p_{i}p^{i}
\label{5}
\end{equation}
As in GR chaos case the system considered here has several conserved quantities. In the non-Riemannian case , besides the magnitude of mass m and the spin S , which is given by
\begin{equation}
S^{2}:= \frac{1}{2}S_{ij}S^{ij}
\label{6}
\end{equation}
which are constants of motion S. Hojman \cite{7} has derived the non-Riemannian constant of motion in spacetimes with torsion which reads 
\begin{equation}
C:= {\xi}^{j}p_{j}-\frac{1}{2}[{\xi}_{{j},{k}}- K_{ljk}{\xi}^{l}] S^{jk}
\label{7}
\end{equation}
where the coma denotes partial derivative. In the next section we shall consider the angular-momentum constants from the Killing constants C and the constraints equations. Here ${\xi}^{i}$ is the Killing vector obeying the well-known Killing equation 
\begin{equation}
{\xi}_{(j;k)}=0
\label{8}
\end{equation}
and the Lie derivative of the contortion tensor
\begin{equation}
{\L}_{\xi}K_{ijk}=0
\label{9}
\end{equation}
where the semi-collon denotes the Riemannian covariant derivative. In our case we shall not be using the Lie derivative of contortion since the exterior metric of the spinning string as we shall see bellow iis on a spacetime which does not contain torsion off the spacetime defect. Let us consider the exterior metric to the spinning string in EC gravity given by Soleng
\begin{equation}
ds^{2}=-dt^{2}+dr^{2}-\frac{{\sigma}{\mu}}{{\lambda}{\pi}}dtd{\phi}+[(1-\frac{\mu}{2{\pi}})^{2}(r+r_{0})^{2}-\frac{{\sigma}^{2}{\mu}^{2}}{4{\lambda}^{2}{\pi}^{2}}]d{\phi}^{2}+dz^{2}
\label{10}
\end{equation}
where ${\sigma}$ is the constant spin density of the spin polarized spinning string and ${\mu}$ is the mass density of the string while $r_{0}$ is the string radius. The only non-vanishing Cartan contortion components are given by expression
\begin{equation}
K_{tr{\phi}}=-\frac{1}{2}T_{tr{\phi}}
\label{11}
\end{equation}
\begin{equation}
K_{{\phi}r{\phi}}= -T_{{\phi}r{\phi}}
\label{12}
\end{equation}
where the only non-vanishing components 
along with the torsion components
\begin{equation}
{T^{t}}_{r{\phi}}= {\sigma}
\label{13}
\end{equation}
where we have computed the contortion components through the expression 
\begin{equation}
K_{ijk}= \frac{1}{2}(T_{jik}+ T_{kij}- T_{kij})
\label{14}
\end{equation}
Let us now consider the  three Killing vectors 
\begin{equation}
{\xi}^{i}=(1,0,0,0)
\label{15}
\end{equation}
\begin{equation}
{\eta}^{i}=(0,0,1,0)
\label{16}
\end{equation}
\begin{equation}
{\epsilon}^{i}=(0,0,0,1)
\label{17}
\end{equation}
which covariant components read 
\begin{equation}
{\xi}_{j}=(-1,0,-\frac{j}{\pi},0)
\label{18}
\end{equation}
\begin{equation}
{\eta}_{j}=(0,0,[1-\frac{\mu}{2{\pi}}]^{2}(r+r_{0})^{2},0)
\label{19}
\end{equation}
\begin{equation}
{\epsilon}_{j}=(0,0,0,1)
\label{20}
\end{equation}
Substitution of these Killing vectors into the Hojman conserved quantities, one obtains
\begin{equation}
E= -p_{t}
\label{21}
\end{equation}
which is the orbit energy which does not explicitly displays the spin-contortion interaction for the orbit which physically means that we are not in a spacetime with torsion off the string. If one writes the energy expression in terms of the contravariant components of the momenta one obtains
\begin{equation}
E= p^{t}+ \frac{{\sigma}{\mu}}{2{\pi}{\lambda}}
\label{22}
\end{equation}
which reveals the presence of the spin density of the spin polarized spinning string. 
\begin{equation}
J_{z}= p_{z}
\label{23}
\end{equation}
\begin{equation}
J_{r}= p_{r}
\label{24}
\end{equation}
\begin{equation}
J_{\phi}= C_{2}= p_{\phi} -[(1-\frac{\mu}{2{\pi}})]^{2}S^{{\phi}r} 
\label{25}
\end{equation}
where to perform this computation use has been made of the following nonvanishing components of the spin tensor $S^{ij}$ as
\begin{equation}
s= (S^{tr},S^{t{\theta}},S^{t{\phi}},S^{{\theta}{\phi}},S^{{\phi}r},S^{r{\theta}})
\label{26}
\end{equation}
Let us now choose the direction of the $3$-vector of angular momentum $(J_{r},J_{\phi},J_{z})$ as 
\begin{equation}
(p_{r},J,0)
\label{27}
\end{equation}
with this choice the previous relation of the angular momentum component $J_{\phi}$ reads
\begin{equation}
S^{{\phi}r}= \frac{[p_{\phi}-J]}{2r_{0}(1-\frac{\mu}{2{\pi}})^{2}}
\label{28}
\end{equation}
To try to determine the remaining components of the spin tensor of the spinning test particle we have to make use of our choice of $p_{z}=0$, which means that we  only deal with planar motion of spinning particles in this paper, and the center of mass relation reduce to the following algebraic relations 
\begin{equation}
-E S^{zt}+S^{z{\phi}}p_{\phi} + S^{zr}p_{r}=0
\label{29}
\end{equation}
\begin{equation}
S^{tr}p_{r}+S^{t{\phi}}p_{\phi}=0
\label{30}
\end{equation}
\begin{equation}
S^{{\phi}r}p_{r}-ES^{{\phi}t}=0 
\label{31}
\end{equation}
Note that from these expressions one is able to obtain an expression for the total energy E in terms of the linear momenta and the spin tensor 
\begin{equation}
E=\frac{1}{S^{zt}}[p_{\phi}S^{z{\phi}}+S^{zr}p_{r}] 
\label{32}
\end{equation}
unfortunatly only the algebraic constraints on the spin tensor are not sufficient to determine the energy since in this expression only $p_{\phi}$ is determined. In the next section I shall compute the energy expressions for the open and closed planar orbits which will yield continuos and discrete spectra respectively.
\section{Continuos Spectra and Energy Shift for Planar Orbits}
In this section we shall make use of last  expression of the previous section to compute the energy spectra of an open planar orbit which is a continuos spectra. This can be done by substitution expression (\ref{28}) into (\ref{32}) which yields
\begin{equation}
E= {\alpha}p_{r}+{\beta}[(1-\frac{\mu}{2{\pi}})^{2}(r+r_{0})S^{{\phi}r}-J]
\label{33}
\end{equation}
where ${\alpha}=\frac{S^{zr}}{S^{zt}}$ and ${\beta}=\frac{S^{z{\phi}}}{S^{zt}}$. The ratio $\frac{\alpha}{\beta}$ gives us the relative spin direction and maybe useful in investigated the chaotic behaviour of the spinning particles around the cosmic spinning string. Note that even when the angular momentum J vanishes or the spin vanishes the spectrum is still continuos or there is no splitting on the energy spectrum by the spin of the particles around the open orbit. This was expected since the particle is not strongly bound to the system and can scape away from the spinning string. To obtain the discrete spectrum we now make use of the relation (\ref{5}) which can explicitly written as
\begin{equation}
m^{2}=-p_{i}p^{i}= -(p_{t}p^{t}+p_{r}p^{r}+p_{\phi}p^{\phi})
\label{34}
\end{equation}
Since the orbit is planar $p_{z}=p^{z}=0$ and circular $p^{r}=0$ and $r=r_{0}$ expression (\ref{34} simplifies to
\begin{equation}
m^{2}=-p_{i}p^{i}= -(p_{t}p^{t}+p_{\phi}p^{\phi})=[(p^{t})^{2}-2g_{t{\phi}}p^{\phi}p^{t}g_{{\phi}{\phi}}(p^{\phi})^{2}]
\label{35}
\end{equation}
The chice $r=r_{0}$ of tangential circular motion is not of course necessary but we shall use it for simplification purposes. This expression  allows us to a second order algebraic equation for the total energy E. To this aim we shall make use of the contravariant components of the metric as 
\begin{equation}
g^{tt}= \frac{{\pi}^{2}b^{2}-\frac{1}{4}j^{2}}{\frac{3}{4}j^{2}-b^{2}}
\label{36}
\end{equation}
\begin{equation}
g^{t{\phi}}= \frac{\frac{j}{\pi}}{\frac{3}{4}j^{2}+b^{2}{\pi}^{2}}
\label{37}
\end{equation}
\begin{equation}
g^{rr}=g^{zz}= 1
\label{38}
\end{equation}
\begin{equation}
g^{{\phi}{\phi}}= \frac{{\pi}^{2}}{\frac{3}{4}j^{2}-b^{2}{\pi}^{2}}
\label{39}
\end{equation}
where $b^{2}=2(1-\frac{\mu}{2{\pi}})^{2}r_{0}$. With these expression for the contravariant components of the metric in hand one may transform the energy second-order algebraic equation 
\begin{equation}
(1-2g^{t{\phi}}g^{t{\phi}})E^{2}+ 3 g_{t{\phi}}g^{{\phi}{\phi}}p_{\phi}E-(m^{2}+g_{{\phi}{\phi}}p_{\phi}g^{{\phi}{\phi}})=0 
\label{40}
\end{equation}
This equation contains two distinct solutions which are responsible for the upper $(E_{+})$ and lower bound $(E_{-})$ for the energy splitting of the energy spectrum, therefore the energy shift can now be computed from the formula
\begin{equation}
{\Delta}E= E_{+} - E_{-}
\label{41}
\end{equation}
which from (\ref{40}) and expressions (\ref{37}),(\ref{39}) after some algebra yields
\begin{equation}
{\Delta}E=\frac{\sqrt{\frac{9j^{2}}{[\frac{3}{4}j^{2}-b^{2}{\pi}^{2}]^{2}}(J+b^{2}S^{{\phi}r})-4{(1-2\frac{j^{2}{\pi}^{2}}{[\frac{3}{4}j^{2}+b^{2}{\pi}^{2}]})[m^{2}+(J+S^{{\phi}r}}b^{2})]\frac{b^{2}{\pi}^{2}-j^{2}}{\frac{3}{4}j^{2}-b^{2}{\pi}^{2}}}}{[1-2g^{t{\phi}}]} 
\label{42}
\end{equation}
One may note that the $S^{r{\phi}}j^{2}$ term represents the spin-spin effect contribution to the energy shift between the spin of the particle and the spin of the string, while the term $J j^{2}$ represents the spin-angular momentum relation between the angular momentum of the string and the total angular momentum J . This term can be considered as a spin-orbit term. Note besides that if the spin tensor component $S^{{\phi}r}>0$ the spin-spin effect contributes negatively to the energy shift while a positive spin tensor component contributes so as to increase the energy shift.This would correspond to the repulsive or attractive nature of the spin-spin interaction. Note that for spinless particles $(S^{r{\phi}}=0)$ and angular momentum $J=0$ different from zero the energy shift of spectrum 
\begin{equation}
{\Delta}E=\frac{\sqrt{\frac{9j^{2}}{[\frac{3}{4}j^{2}-b^{2}{\pi}^{2}]^{2}}-4{(1-2\frac{j^{2}{\pi}^{2}}{[\frac{3}{4}j^{2}+b^{2}{\pi}^{2}]})[m^{2}]}}}{[1-2g^{t{\phi}}]} 
\label{43}
\end{equation}
Similar results for the nonclassical Dirac particles including the geodesic classical limit also discussed here for spinning strings were obtained by Figueiredo et al \cite{2,3} in the case of G\"{o}del metrics.
\section{Discussions}
There are several interesting prospects for the future in the research we reported here. The first step would be to generalise the problem discussed here to include gravitational waves in the background of the spinning string spacetime and try to solve the equations of motion for the spinning particle. Another intersting project would be to extend the work discussed here for another types of torsion defects such as planar walls and Superconducting cosmic strings \cite{14}. The spin-spin interaction investigated is also similar to the more recently interaction found by Bonnor \cite{15} between two spinning strings.
 
\section*{Acknowledgement}
I am deeply grateful to Professors I.D.Soares and P. S. Letelier for helpful discussions on the subject of this paper. Thanks are also due to Professor W. Bonnor and Dr. C. N. Ferreira for her interest in this project. Financial support from CNPq. and UERJ is gratefully ackowledged.

\end{document}